\documentclass[prd,twocolumn,eqsecnum,nofootinbib]{revtex4}
\usepackage{bm}
\usepackage{amssymb}
\usepackage{graphicx}
\usepackage{epstopdf}
\usepackage{mathrsfs}
\usepackage{amsmath}

\begin{document}
\title{The Marolf-Ori singularity inside fast spinning black holes} 
\author{Lior M.~Burko$^1$ and Gaurav Khanna$^2$}
\affiliation{
$^1$ School of Science and Technology, Georgia Gwinnett College, Lawrenceville, Georgia 30043 \\ 
$^2$ Department of Physics, University of Massachusetts, Dartmouth, Massachusetts  02747}
\date{January 10, 2019}
\begin{abstract} 

The effective shock wave singularity at the outgoing leg of the inner horizon of a linearly perturbed fast spinning black hole is studied numerically for either scalar field, or vacuum, gravitational perturbations. We demonstrate the occurrence of the Marolf-Ori singularity, including changes of order unity in the scalar field $\phi$ for the scalar field model, and in the Weyl scalars $\psi_0$ and $\psi_4$ (rescaled appropriately by the horizon function $\Delta$) and the Kretschmann curvature scalar $K$ for the vacuum, gravitational perturbations model for both null and timelike geodesic observers. We quantify the shock sharpening effect and show that in all cases its rate agrees with expectations.

\end{abstract}
\maketitle

The fate of an astronaut who falls into a black hole depends not just on the latter's properties (such as the intrinsic parameters, i.e., the mass and  spin angular momentum, and the external perturbation fields) but also on the former's worldline. Specifically, for geodesic equatorial timelike geodesics as mapped on the spacetime of the corresponding unperturbed Kerr black hole, astronauts with positive energy and high values of their angular momentum generally end up at a null, weak singularity at the ingoing leg of the black hole's inner horizon, the Cauchy horizon (CH) singularity (``mass inflation singularity", ``infalling singularity") \cite{Ori-92}. However, astronauts with positive energy and low angular momentum (including counterrotating ones) arrive at the outgoing leg of the black hole's inner horizon (``outgoing inner horizon", henceforth, OIH). 

The properties of spacetime at the OIH have been proposed to be those of an effective shock wave singularity \cite{Marolf-Ori-12}. Specifically, it was proposed in \cite{Marolf-Ori-12} that daughters of a family of  free-falling astronauts whose geodesics intersect with the OIH, and who are separated only by time translations (and labeled by the advanced time values at which they cross the event horizon (EH), $v_{\rm eh}$) experience a change of order unity in typical metric perturbations, and that these changes occur over a lapse of proper time that drops like 
$\sim e^{-\kappa v_{\rm eh}}$ with increasing $v_{\rm eh}$, where $\kappa$ is the surface gravity of the OIH. Sufficiently late-falling daughters therefore experience an effective shock wave singularity, the Marolf-Ori singularity (``outflying singularity"). 

\begin{figure}[t!]
\includegraphics[width=6.5cm]{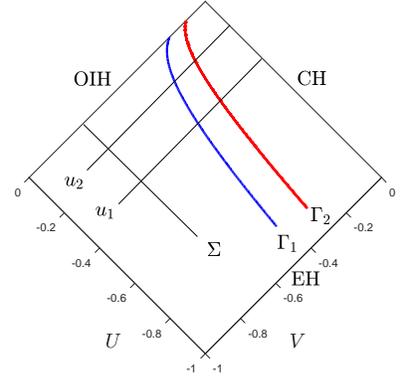}
\caption{Penrose diagram of the simulated spacetime, shown in compactified Kruskal-like coordinates $U,V$. The EH is at $U=-1$, the CH is at $V=0$ and the OIH is at $U=0$. Two timelike geodesics ($\Gamma_1,\Gamma_2$) are shown, in addition to two outgoing null rays ($u_1,u_2$) and an ingoing null ray ($\Sigma$).}
\label{penrose}
\end{figure}

\begin{figure}[t!]
\includegraphics[width=6.5cm]{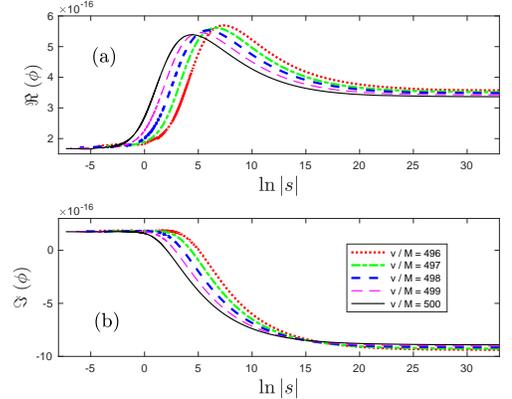}
\caption{The scalar field $\phi$ as a function of $s$ for a family of ingoing null geodesics. Upper panel: the real part, $\Re (\phi)$.  Lower panel: the imaginary part, $\Im (\phi)$. In each case the field is shown for five geodesics, the earliest of which is at $v/M=496$ in increments of $\,\Delta v=M$.}
\label{null_scalar}
\end{figure}

The Marolf-Ori singularity evolves because incoming radiation which travels along an ingoing null ray in the past of the infalling observers (see Fig.~\ref{penrose}) and is scattered outward by spacetime curvature, is observed differently by late daughters than by earlier daughters: As function of retarded time $u$, the radiation pattern between two outgoing null rays is little changed between different daughters. But as the proper time difference along the daughter's worldline between these two outgoing null rays  behaves like $\sim e^{-\kappa v_{\rm eh}}$ (or, equivalently, as the relativistic $\gamma$-factor increases exponentially, say with respect to some natural frame near the OIH), any feature in the radiation field is sharpened exponentially fast with $v_{\rm eh}$. Specifically, the order unity changes in a scalar field, or in the Weyl scalars $\psi_0$ and $\psi_4$  become effective shock waves when the time scale for the change in the fields becomes too small for a physical observer to measure.

When fully nonlinear evolution of the interior spacetime is considered, there may be a third class of astronauts, who intersect the singularity at a spacelike sector, where the singularity may be of the Belinskii, Khalatnikov, and Lifshitz (BKL) type \cite{BKL}. Evidence for a spacelike (and strong) singularity inside perturbed black holes is currently available only for the spherical charged toy model, but without the chaotic BKL behavior \cite{burko-97}. We do not consider here the possible occurrence of a spacelike singularity inside rotating black holes. We also do not consider the effect of accretion of baryons or dark matter \cite{hamilton} or that of absorbed photons from the cosmic background radiation \cite{burko-97b}, or black holes that are asymptotically de Sitter. The perturbation fields considered in this paper are those of an asymptotically-flat, isolated Kerr black hole perturbed (linearly) by scalar fields or gravitational waves that result from the Price tails \cite{price} that follow the collapse \cite{burko-khanna-2014}.

The evidence beyond the original work \cite{Marolf-Ori-12} for the occurrence of the Marolf-Ori singularity  has focused mostly on the toy model of a spherical charged black hole with scalar fields \cite{eilon-ori}, neutral null fluids \cite{eilon} or a combination of the two \cite{eilon}. The Marolf-Ori singularity was also found for rotating black holes with fully nonlinear scalar fields, for a model that considered initial data posed in the interior of the black hole \cite{chesler}. It is as yet unclear how such initial data can arise from evolutionary processes of generic external or internal perturbations. The occurrence of the Marolf-Ori singularity for the model considered in \cite{chesler} is strong evidence for the robustness of the shock wave singularity.

\begin{center}
\begin{table}
\begin{tabular}{ || l | c | c || }
  \hline			
  Quantity & $\alpha$ for Real part & $\alpha$ for Imaginary  part \\
  \hline
 $\Delta^2\psi_0$ & $-1.0011\pm 0.0007$ & $-1.0033\pm 0.0016$ \\
 $\Delta^{-2}\psi_4$ & $-0.994\pm 0.021$ & $-1.028\pm 0.040$ \\
 $K$ & $-0.963\pm 0.046$ & --- \\
 \hline
 $\phi$ & $-0.989\pm 0.058$ & $-1.008\pm 0.004$ \\
  \hline  
  \hline
   $\Delta^2\psi_0$ & $-1.0010\pm 0.00013$ & $-0.999\pm 0.0025$ \\
 $\Delta^{-2}\psi_4$ & $-1.035\pm 0.019$ & $-1.047\pm 0.010$ \\
 $K$ & $-1.036\pm 0.009$ & --- \\
 \hline
 $\phi$ & $-1.000\pm 0.015$ & $-0.999\pm 0.0025$ \\
 \hline
\end{tabular}
\caption{The parameter $\alpha$ for null geodesics (above the horizontal double lines) and ZAMOs (below the horizontal double lines)  intersecting with the OIH. }
\label{tab1}
\end{table}
\end{center}

\begin{figure}[t!]
\includegraphics[width=6.5cm]{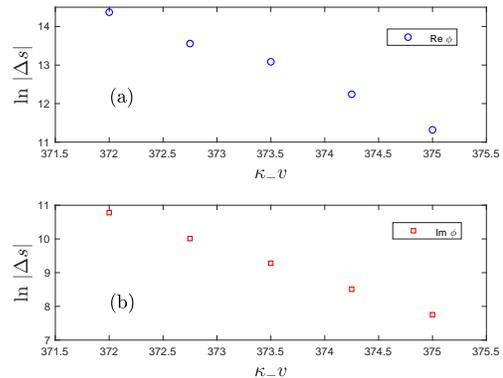}
\caption{The (natural logarithm of the) difference $\,\Delta s$ as a function of $\kappa v$ for the real (upper panel) and imaginary (lower panel) parts of the scalar field $\phi$. The slope of each curve is denoted by $\alpha$.}
\label{null_scalar_sharpening}
\end{figure}

\begin{figure}[h!]
\includegraphics[width=7.5cm]{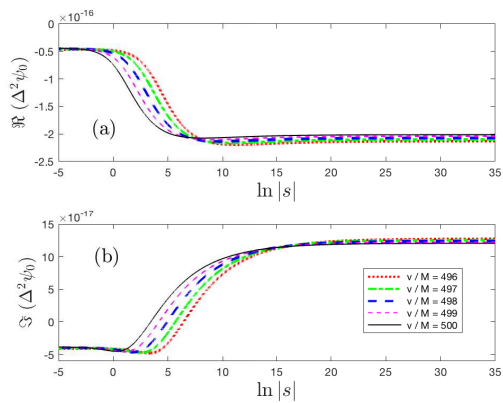}
\caption{The Weyl scalar $\psi_0$ (multiplied by $\Delta^2$) as a function of $s$ for a family of ingoing null geodesics. Upper panel: the real part, $\Re (\Delta^2\phi)$.  Lower panel: the imaginary part, $\Im (\Delta^2\phi)$. In each case the field is shown for five geodesics, the earliest of which is at $v/M=496$ in increments of $\,\Delta v=M$.}
\label{psi0}
\end{figure}

Here, we consider in detail for the first time the occurrence and properties of the Marolf-Ori singularity for vacuum, gravitational  perturbations (we also consider scalar fields) inside fast spinning black holes for astrophysically realistic initial data, within the linear approximation. This approximation allows us to find the behavior of the $\psi_0$ and $\psi_4$ Weyl scalars and the behavior of the Kretschmann curvature scalar $K$ (or the scalar field $\phi$ itself). It does not allow us, however, to find inherently nonlinear effects such as the behavior of metric functions.

\begin{figure}[t!]
\includegraphics[width=7.5cm]{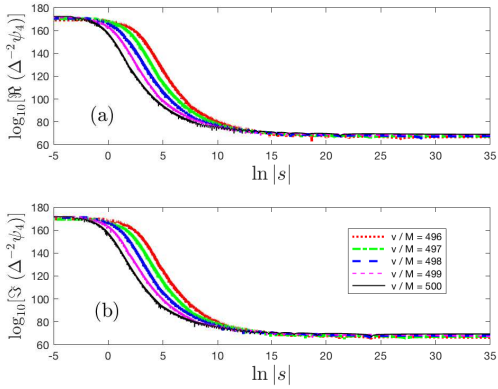}
\caption{Same as Fig.~\ref{psi0} for the Weyl scalar $\psi_4$ (multiplied by $\Delta^{-2}$). The horizon function $\Delta=(r_+-r)(r-r_-)$, where $r_{\pm}$ are the values of the $r$ coordinate at the outer and inner horizons, correspondingly.}
\label{psi4}
\end{figure}

\begin{figure}[h!]
\includegraphics[width=7.5cm]{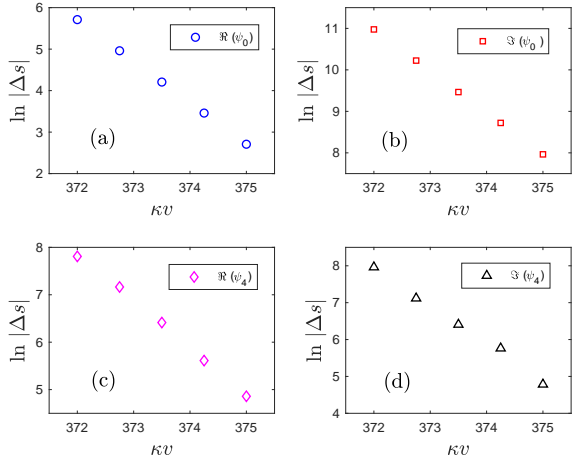}
\caption{Same as Fig.~\ref{null_scalar_sharpening} for the real and imaginary parts of  $\psi_0$ (multiplied by $\Delta^{2}$, upper panels) and of $\psi_4$ (multiplied by $\Delta^{-2}$, lower panels).}
\label{psi0_sharp}
\end{figure}

We present numerical results from the solution of the 2+1 dimensional Teukolsky equation for a Kerr black hole with mass $M=1$ and spin angular momentum $a=0.8M$, using the methods described in \cite{BKZ}. Initial data for any of the fields we show are chosen to be truncated Gaussians centered at $\rho=5.0$ with width of 0.2, vanishing outside the domain $3<\rho<7$. Here, $\rho$ is the compactified coordinate defined in \cite{BKZ}. 

First, we consider ingoing null observers, parametrized by their value of advanced time (``Eddington coordinate") $v$. 
The real and imaginary parts of the scalar field $\phi$ are shown in Fig.~\ref{null_scalar} as functions of $s= -e^{-\kappa(u+v)}$. Near the OIH, $s$ is a good approximation for the affine parameter. 
Figure \ref{null_scalar} shows the order unity change in $\phi$, which does not change appreciably for later null geodesic.

\begin{figure}[h]
\includegraphics[width=6.5cm]{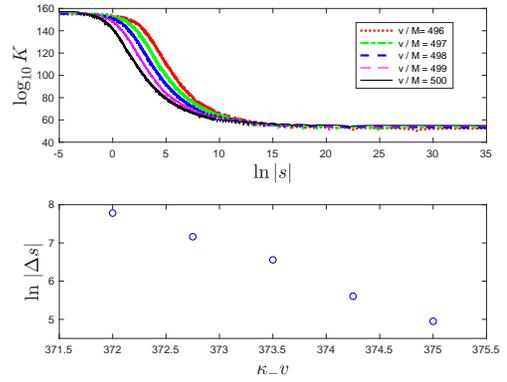}
\caption{Upper panel: Same as Fig.~\ref{psi0} for the curvature scalar $K$. Lower panel: Same as Fig.~\ref{null_scalar_sharpening} for the curvature scalar $K$.}
\label{k-null}
\end{figure}

\begin{figure}[h]
\includegraphics[width=7.5cm]{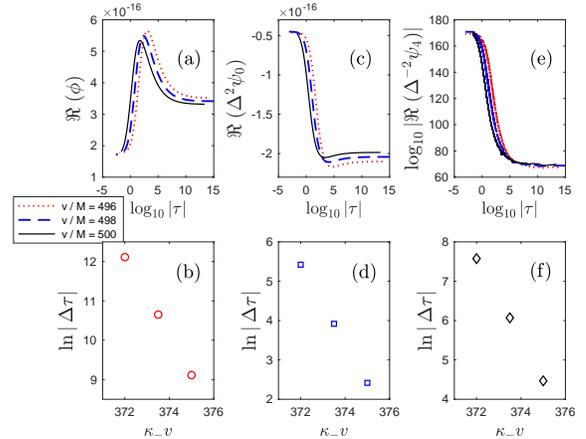}
\caption{Upper three panels: the real parts of $\phi$, $\,\Delta^2\psi_0$, and $\,\Delta^{-2}\psi_4$ as functions of proper time $\tau$ (expressed in units of the black hole mass $M$). Lower three panels: The changes $\,\Delta\tau$ as functions of $\kappa v_{\rm eh}$ corresponding to the three cases of the upper panel. }
\label{timelike_real}
\end{figure}

\begin{figure}[h!]
\includegraphics[width=7.5cm]{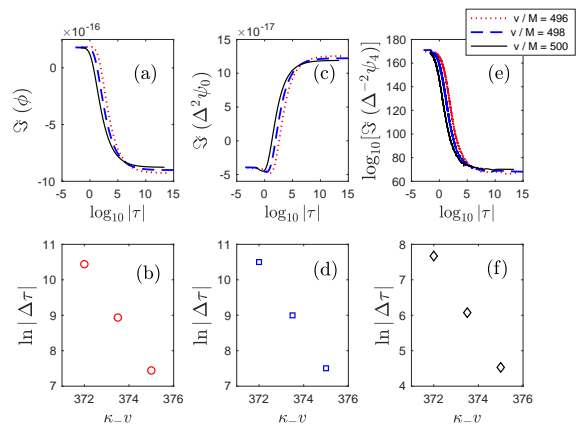}
\caption{Same as Fig.~\ref{timelike_real} for the imaginary parts of  $\phi$, $\,\Delta^2\psi_0$, and $\,\Delta^{-2}\psi_4$.}
\label{timelike_imaginary}
\end{figure}

\begin{figure}[h!]
\includegraphics[width=6.5cm]{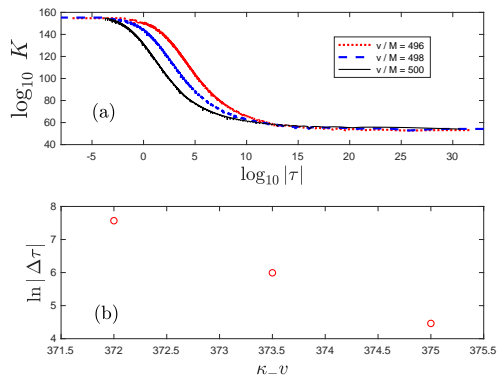}
\caption{Same as Fig.~\ref{k-null} for the timelike case. Here, proper time $\tau$ is used instead of $s$.}
\label{k-timelike}
\end{figure}

The shock sharpening effect is evident using the qualitative argument used in \cite{eilon-ori}: As the width of each curve in Fig.~\ref{null_scalar}, $\,\Delta (\ln\, | s |)$, is roughly the same for all these geodesics, the smaller the values of $\ln\, | s |$ the narrower the width. The shock sharpening effect can be demonstrated quantitatively by finding the width of the change in $\phi$. For $\Re (\phi)$ we determine the width by finding the difference in $s$ between  the values of $s$ at which $\phi$ equals $75\%$ its peak value above its minimum. For $\Im (\phi)$ we determine the width by finding $\,\Delta s$ between $25\%$ and $75\%$ of the change in $\phi$. We then plot in Fig.~\ref{null_scalar_sharpening} $\ln \,\Delta s$ for each null geodesic as a function of $\kappa v$ for the real and imaginary cases. We denote the slope of each curve by the parameter $\alpha$. Based on the analysis of \cite{Marolf-Ori-12,eilon-ori} we expect $\alpha=-1$. The values we measure appear in Table \ref{tab1}. For both the real and the imaginary parts of $\phi$ we find agreement between the predicted value of $\alpha$ and its measured value.

\begin{figure}[h]
\includegraphics[width=6.5cm]{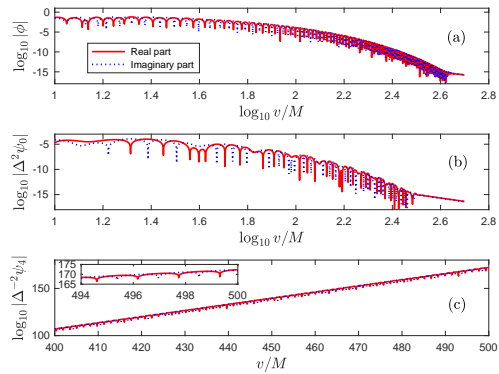}
\caption{The real and imaginary parts of field along the OIH as functions of $v/M$. Top panel: The scalar field $\psi$; Middle panel: $\,\Delta^2\psi_0$; Lower panel:  $\,\Delta^{-2}\psi_4$.}
\label{oih}
\end{figure}

Next, we consider gravitational perturbations. Figures \ref{psi0} and \ref{psi4} show (the real and imaginary parts of) the Weyl scalars $\psi_0$ and $\psi_4$ in the Hartle-Hawking tetrad \cite{hartle-hawking-1972}, respectively. The width for each case is determined as for $\Im (\phi)$. Figure \ref{psi0_sharp} shows the shock sharpening effect for $\psi_0$ and $\psi_4$. Notice, that both $\,\Delta^2\psi_0$ and $\,\Delta^{-2}\psi_4$ experience an order unity change in magnitude, and approach finite values as $s\to\ 0$. Therefore, $\psi_0\to\infty$ and $\psi_4\to 0$ as $s\to\ 0$. However, as the Weyl scalars transform under tetrad rotation, the latter conclusion is not tetrad independent. A quantity which is tetrad independent is the Kretschmann scalar $K\sim 8\psi_0\psi_4+{\rm c.c}$. (Note, that the Hartle-Hawking tetrad is a transverse frame, i.e., $\psi_1=0=\psi_3$, and that $\psi_2$ is that of the background Kerr spacetime.) Figure \ref{k-null} shows the behavior of $K$ and the respective shock sharpening effect. 
For all three cases of $\psi_0, \psi_4$, and $K$ we find the parameter $\alpha$ to be in agreement with the expected value (see Table \ref{tab1}). 

Timelike geodesics are chosen to be a family of geodesic observers with energy $E/\mu=1$ and zero angular momentum (ZAMOs)  $L/\mu=0 M$, which are separated only by $v_{\rm eh}$. Here, $\mu$ is the mass of the freely falling observer.  These geodesics intersect with the OIH. (For $E>0$, the condition that a timelike geodesic intersect with the CH and not with the OIH is that $L>2EMr_-/a$.) Figures \ref{timelike_real} and \ref{timelike_imaginary}  show the real and imaginary parts, respectively, of $\phi$, $\,\Delta^2\psi_0$, and $\,\Delta^{-2}\psi_4$ as functions of proper time $\tau$, and the behavior of $\,\Delta\tau$ for each geodesic as a function of $\kappa v_{\rm eh}$. Here, $\tau=0$ when the geodesic intersects with the OIH. 
Figure \ref{k-timelike} shows the same for the Kretschmann scalar $K$. 
The widths of the changes in the fields' values are determined as above. 
In all cases we find the parameter $\alpha$ to be in agreement with the expected value. 

The effective shock wave singularity is related to the transverse direction (i.e., the direction of $\,\partial / \,\partial s$ ($\,\partial / \,\partial \tau$) for the null (timelike) case). In the direction along the OIH ($\,\partial / \,\partial v$) the fields behave as along any other outgoing null geodesic that intersects with the CH \cite{BKZ}. Figure \ref{oih} shows the (real and imaginary parts of the) scalar field, $\,\Delta^2\psi_0$ and $\,\Delta^{-2}\psi_4$ as functions of $v$ along the OIH. 

We have shown evidence for the evolution of a Marolf-Ori singularity for vacuum perturbations, and  for its evolution inside rotating black holes that are perturbed by external perturbations. 
It is as yet an open question whether the OIH survives (even as a Marolf-Ori singularity) when the black hole is formed in a  fully nonlinear dynamical collapse process. Even if it does, the question of the fate of an astronaut whose worldline intersects with the OIH awaits further consideration. It is conceivable that the deformation of a physical object may be approximated by a step response, which suggests that it would oscillate about some deformed state, and the magnitude of the new equilibrium deformation may be comparable to the object's original dimensions. It is yet to be assessed how the internal structure of a physical object would respond to such strains. 

The authors thank Amos Ori for discussions and useful comments made on an earlier version of this paper. 
G.\ K.\ acknowledges research support from NSF Grant PHY-1701284 and ONR/DURIP Grant No.\ N00014181255.

\end{document}